\documentclass[letter,oldversion]{aa}
\usepackage{graphicx}
\usepackage{txfonts}
\usepackage{natbib}
\usepackage{color}
\usepackage{url}
\usepackage{array,multirow}

\begin{document}

   \title{The Gaia-ESO Survey: \\ A globular cluster escapee in the Galactic halo \thanks{Based on data acquired by the Gaia-ESO Survey, programme ID 188.B-3002. Observations were made with ESO Telescopes at the La Silla Paranal Observatory.}}

   \author{	K. Lind\inst{1,2}\and
	        	S.~E. Koposov\inst{2,3}\and
	        	C. Battistini\inst{4}\and
   	        	A.~F. Marino\inst{5}\and
	        	G. Ruchti\inst{4}\and
	        	A. Serenelli\inst{6}\and
	        	C.~C. Worley\inst{2}\and
 		A. Alves-Brito\inst{5}\and 
 		M. Asplund\inst{5}\and 
 		P.~S. Barklem\inst{1}\and 
 		T. Bensby\inst{4}\and 
 		M. Bergemann\inst{2}\and
 		S. Blanco-Cuaresma\inst{7}\and
		A. Bragaglia\inst{11}\and
 		B. Edvardsson\inst{1}\and 
 		S. Feltzing\inst{4}\and
 		P. Gruyters\inst{1}\and
 		U. Heiter\inst{1}\and 
 		P. Jofre\inst{2}\and
 		A.~J. Korn\inst{1}\and 
 		T. Nordlander\inst{1}\and
 		N. Ryde\inst{4}\and 
 		C. Soubiran\inst{7}\and
		G. Gilmore\inst{2}\and
		S. Randich\inst{8}\and	
		A.~M.~N. Ferguson\inst{18}\and         
		R.~D. Jeffries\inst{17}\and
		A. Vallenari\inst{16}\and	
		C. Allende Prieto\inst{9,10}\and
		E. Pancino\inst{11,12}\and		
		A. Recio-Blanco\inst{13}\and
		D. Romano\inst{11}\and
		R. Smiljanic\inst{14}\and
		M. Bellazzini\inst{11}\and
		F. Damiani\inst{19}\and
		V. Hill\inst{13}\and
		P. de Laverny\inst{13}\and
		R.~J. Jackson\inst{17}\and
		C. Lardo\inst{15}\and
		S. Zaggia\inst{16}
           }

   \institute{
             Department of Physics and Astronomy, Uppsala University, Box 516, 75120, Uppsala, Sweden\\                 
   	    \email{karin.lind@physics.uu.se}\and
             Institute of Astronomy, University of Cambridge, Madingley Road, Cambridge, CB3 0HA, United Kingdom\and                  
             Moscow MV Lomonosov State University, Sternberg Astronomical Institute, Moscow 119992, Russia\and
             Lund Observatory, Department of Astronomy and Theoretical Physics, Box 43, SE-221 00 Lund, Sweden\and                 
             Research School of Astronomy \& Astrophysics, Australian National University, Cotter Road, Weston Creek, ACT 2611, Australia\and                 
             Institute of Space Sciences (IEEC-CSIC), Campus UAB, Fac. Ci\'{e}ncies, Torre C5 parell 2, 08193 Bellaterra, Spain\and                 
             Lab. d'Astrophysique de Bordeaux, CNRS, Universit\'{e} Bordeaux, 2 rue de l'Observatoire, BP 89, 33271 Floirac Cedex, France\and                  
             INAF - Osservatorio Astrofisico di Arcetri, Largo E. Fermi 5, 50125, Florence, Italy\and                 
             Instituto de Astrof\'{\i}sica de Canarias, E-38205 La Laguna, Tenerife, Spain\and
             Universidad de La Laguna, Dept. Astrof\'{\i}sica, E-38206 La Laguna, Tenerife, Spain\and
             INAF - Osservatorio Astronomico di Bologna, via Ranzani 1, 40127, Bologna, Italy\and	
             ASI Science Data Center, Via del Politecnico SNC, 00133 Roma, Italy\and 
	    Lab. Lagrange, Universit\'e de Nice Sophia Antipolis, CNRS, Observatoire de la C\^ote d'Azur, BP 4229, 063 04 Nice cedex 4, France\and
             Department for Astrophysics, Nicolaus Copernicus Astronomical Center, ul. Rabia\'{n}ska 8, 87-100 Toru\'{n}, Poland\and
             Astrophysics Research Institute, Liverpool John Moores University, 146 Brownlow Hill, Liverpool L3 5RF, United Kingdom\and	
             INAF - Padova Observatory, Vicolo dell'Osservatorio 5, 35122 Padova, Italy\and 
             Astrophysics Group, Keele University, Keele, Staffordshire ST5 5BG, United Kingdom \and 
             Institute of Astronomy, University of Edinburgh, Blackford Hill, Edinburgh EH9 3HJ, United Kingdom\and
             INAF - Osservatorio Astronomico di Palermo, Piazza del Parlamento 1, 90134, Palermo, Italy
             }

   \date{Received 19 December 2014; accepted 27 January 2015}

\authorrunning{K.Lind et al.}  \titlerunning{A globular cluster escapee in the Galactic halo}

  \abstract{A small fraction of the halo field is made up of stars that share the light element ($Z\le13$) anomalies characteristic of second generation globular cluster (GC) stars. The ejected stars shed light on the formation of the Galactic halo by tracing the dynamical history of the clusters, which are believed to have once been more massive. Some of these ejected stars are expected to show strong Al enhancement at the expense of shortage of Mg, but until now no such star has been found. We search for outliers in the Mg and Al abundances of the few hundreds of halo field stars observed in the first eighteen months of the Gaia-ESO public spectroscopic survey. 
   One halo star at the base of the red giant branch, here referred to as 22593757-4648029 is found to have $\rm[Mg/Fe]=-0.36\pm0.04$ and $\rm[Al/Fe]=0.99\pm0.08$, which is compatible with the most extreme ratios detected in GCs so far. We compare the orbit of 22593757-4648029 to GCs of similar metallicity and find it unlikely that this star has been tidally stripped with low ejection velocity from any of the clusters. However, both chemical and kinematic arguments render it plausible that the star has been ejected at high velocity from the anomalous GC $\omega$ Centauri within the last few billion years. We cannot rule out other progenitor GCs, because some may have disrupted fully, and the abundance and orbital data are inadequate for many of those that are still intact.}


   \maketitle

\section{Introduction}

The majority of stars in the Galactic halo reside in the field, leaving only 2\% of the total stellar mass in globular clusters \citep[GCs, see e.g.][and references therein]{Vesperini10}. How large a fraction of today's field stars escaped from a cluster and the original sizes of these clusters is a matter of much debate. With few exceptions, field and cluster stars share similar chemical compositions for alpha, Fe-peak and neutron-capture elements \citep{Gratton04,Pritzl05}. However, the GCs display unique abundance patterns for lighter elements that are not typically seen in the field, in particular for C, N, O , Na, Mg, and Al. The existence of such anomalies have been known for decades \citep{Cottrell81,Kraft97} and have been mapped in detail in large GC samples \citep[e.g.][]{Carretta09b}. We now know that multiple populations of stars reside within each cluster, with some stars having been polluted by the elements produced during H-burning at high temperatures by more massive stars \citep{Denisenkov89,Ventura01,Decressin07a}. Consequently, a large fraction of present-day GC stars are enhanced in N and Na, at the cost of depleted levels of C and O. The second generation stars are believed to also be enhanced in He, the main product of H burning. 

The correlating and anti-correlating patterns of the lighter elements characteristic of GCs are seen neither in disk open clusters \citep{deSilva09,Bragaglia14} nor in the Galactic bulge \citep{Lecureur07, Bensby13}. Therefore, by finding out how many field stars have light element anomalies and thus likely originated in GCs, we can study the GC-field link. The largest systematic studies undertaken so far are those by \citet{Martell10} and \citet{Martell11}, who focussed on C and N anomalies as traced by molecular bands in SDSS/Segue spectra. The latter study reports that 3\% of field stars display the chemical characteristics of GC stars. This small number agrees well with higher-resolution studies with fewer number statistics, e.g. \citeauthor[][(2010; 1.4\%)]{Carretta10b} and \citeauthor[][(2012; $3\pm2$\%)]{Ramirez12}. The \citeauthor{Ramirez12} study could confirm both elevated Na and depleted O-abundances in two halo field dwarfs, a clear indication of a GC connection. As discussed, e.g., by \citet{Gratton12}, models of halo formation must account for much higher initial masses of GCs to explain this fraction of second generation GC stars in the field (however, see Prantzos \& Charbonnel 2006 and Bastian et al. 2013).

The O-Na anti-correlation was found in all GCs where it has been investigated with enough statistics, indicating that the progenitor objects were massive enough to activate the NeNa-cycle. However, the anti-correlating abundance pattern of Mg and Al that is expected from the operation of the MgAl-chain manifests itself less uniformly. Several clusters display a small Mg-spread and a significant Al-spread \citep{Carretta09b} while very few clusters, e.g. $\omega$ Cen and NGC\,2808, show a pronounced Mg-Al anti-correlation (see Fig.\,\ref{fig:hist}). The likelihood of finding escaped stars in the field with clear GC characteristics for both Mg and Al is thus small and it requires a large spectroscopic survey to detect them in significant numbers. Here we report on the first discovery of a halo field star with both strongly depleted Mg and strongly elevated Al abundance.

\section{Observations and analysis}
\begin{figure*}
\centering
\includegraphics[width=16.5cm,viewport=1.0cm 12.8cm 28.5cm 20cm,clip]{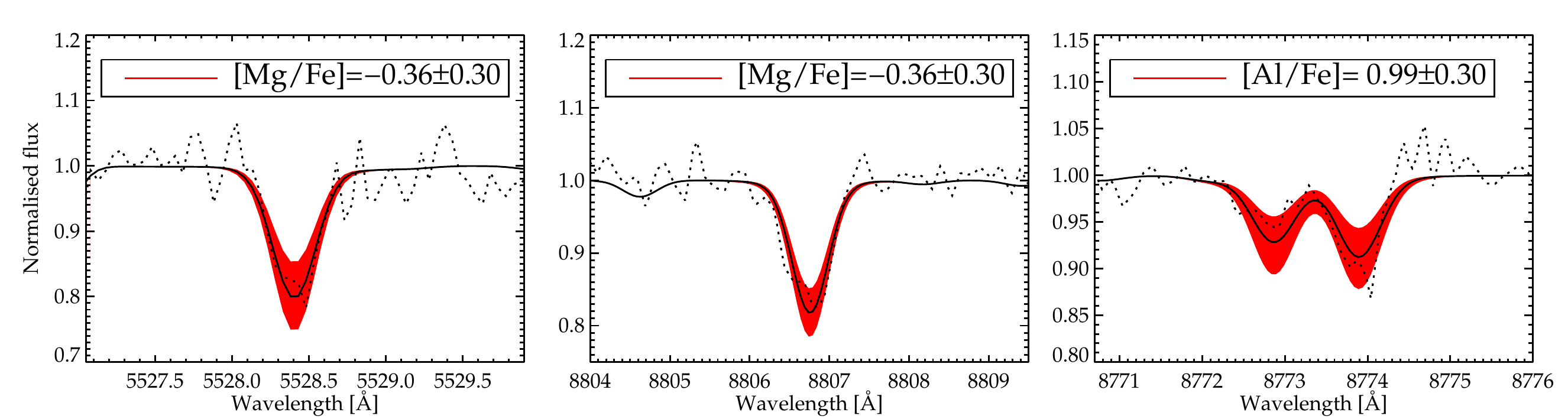}
\caption{Observed \textit{(dotted line)} and best-fit synthetic \textit{(solid line)} spectra of 22593757-4648029. Shaded regions indicate $\pm0.3$\,dex in abundance.}
\label{fig:abund}
\end{figure*}

The Gaia-ESO public spectroscopic survey \citep{Gilmore12} is a five-year survey with FLAMES on the VLT, which has been observing since December 2011. Two multi-object spectrographs operate simultaneously in high resolution (UVES) and medium resolution (GIRAFFE/MEDUSA), with fainter targets such as distant halo stars, allocated to the GIRAFFE fibres. Two settings are used for the halo stars; HR10, which covers 535--565\,nm at a spectral resolving power of $R=\lambda/\delta\lambda=19\,800$, and HR21, which covers 845-900\,nm at $R=16\,200$. The combined analysis of data in the two GIRAFFE settings allows for a determination of stellar parameters, Al, alpha-elements, and Fe-peak elements. For 22593757-4648029, we achieve an average S/N per pixel of 25 in HR10 and 55 in HR21 after 2x1500\,s of exposure in each setting (see sample spectra in Fig.\,\ref{fig:abund}).  The observational data are summarised in Table 1. A full account of the target selection, data reduction and processing will be given in a dedicated GIRAFFE data release paper.

\begin{table}
      \caption{Data for 22593757-4648029.}
         \label{tab:star}
         \centering
         \begin{tabular}{lrlr}
                \hline\hline
                Name  & Value & Name & Value$^{(a)}$\\
 	       \hline
   \noalign{\smallskip} 
                    RA                                	&   22 59 37.57           				& $T_{\rm eff}$ 			& $5261\pm36\pm100$K  \\
                    DEC 	    	           	&  -46 48 02.9      					& $\log{g}$                         	& $2.84\pm0.54\pm0.25$        \\  
                    $V$ 	 			& $15.649\pm0.037$$^{(b)}$  	    		& $\rm[Fe/H]$ 			& $-1.49\pm0.05\pm0.20$    		\\ 
                    $K$ 	 			& $13.928\pm0.061$$^{(c)}$	    		& $\rm[Mg/Fe]$ 		& $-0.36\pm0.04\pm0.10$   		\\ 
                    $J-K$			 	&  $0.494$$^{(c)}$      	    			& $\rm[Al/Fe]$ 			& $0.99\pm0.08\pm0.10$     		\\ 
                    $T_{\rm eff, \it\,J-K}$ 	& $5326\pm140$\,K$^{(d)}$   	    		& $\rm[Si/Fe]$ 			& $0.34\pm0.13\pm0.10$      \\                                                                                                                                                                                                                                                                                                                                                                                                                                                                                                                                                                                                                                                                                                                                                                                                                                                                                                                                                                                                                                                                                                                                                                                                                                                                                                                                                                                                                                                                                                                                                                                                                                                                                                                                                                                                                                                                                                                                                                                                                                                                                                                                                                                                                                                                                                                                                                                                                                                                                                                                                                                                                                                                                                                                                                                                                                                                                                                                                                                  
                    $E_{B-V}$ 			& 0.010$^{(e)}$ 			    		& $\rm[Ca/Fe]$ 		         & $0.41\pm0.13$   		\\
                    $V_{\rm rad}$ 			& $44.97\pm0.16\rm\,km\,s^{-1}$$^{(a)}$  & $\rm[Ti/Fe]$ 			& $0.46\pm0.22$  		\\
                    $\rm \mu_{RA}$ 		& $2.8\pm1.8$\,mas/yr$^{(f)}$  			& $\rm[Y/Fe]$ 			& $<0.11$		\\  
                    $\rm \mu_{DEC}$ 		& $-10.40\pm1.8$\,mas/yr$^{(f)}$  		& Distance			& $6.7^{+1.6}_{-1.3}$\,kpc 		\\ 
   \noalign{\smallskip} 
 	       \hline
   \noalign{\smallskip} 
                    \multicolumn{2}{l}{$^{(a)}$ Gaia-ESO Survey}                                         &  \multicolumn{2}{l}{ $^{(d)}$ \citet{Casagrande10}}\\
                    \multicolumn{2}{l}{$^{(b)}$ APASS }                                                           &  \multicolumn{2}{l}{$^{(e)}$ \citet{Schlegel98}} \\
                     \multicolumn{2}{l}{$^{(c)}$ 2MASS, \citet{Skrutskie06}}                        &  \multicolumn{2}{l}{$^{(f)}$ UCAC } \\                                                    
         \end{tabular}
\end{table}

Several independent analysis nodes, using independent methods, participate in the determination of stellar parameters and chemical abundances in the Gaia-ESO Survey. The results of individual nodes are thoroughly examined, evaluated, and homogenised to arrive at the finally recommended values for each star. We adopt the homogenised, recommended results for 22593757-4648029, based on spectra from the internal data release 2 (GESviDR2Final). The object is classified as a metal-poor star at the base of the red giant branch with $T_{\rm eff}=5250\pm34\pm100$\,K, $\log{g}=2.84\pm0.54\pm0.25$ and $\rm[Fe/H]=-1.49\pm0.05\pm0.20$ (random and systematic error) . The details of all the analysis methods and the homogenisation procedure will be described in Recio-Blanco et al. (in prep). The distance to the star was determined based on Bayesian fits to evolutionary tracks using our derived spectroscopic parameters following the methodology described in \citet{Serenelli13}.

\section{Discussion}

The peculiar chemical composition of 22593757-4648029 is reported in Table \ref{tab:star}. While three of the alpha-elements, Si, Ca, and Ti, are compatible with the standard 0.4\,dex $\rm[\alpha/Fe]$-enhancement characteristic of metal-poor stars, Mg falls approximately 0.8\,dex below the expected value. In contrast, Al is instead enhanced to 1 dex above solar. As discussed below, this is compatible with the most extreme populations found in GCs and we argue that the star is a GC escapee. It was discovered in a sub-sample of $\sim7300$ FGK stars (half of all GIRAFFE targets) with detectable Mg and Al abundances in the disk and halo field. Based on a simple two-component fit to the metallicity distribution function of this sub-sample (peaks at $\rm[Fe/H]=-1.6\pm0.5$ and $\rm[Fe/H]=0.24\pm0.36$) , we expect an approximate halo fraction of 5-10\%, i.e. a few hundred stars. Finding one such chemically unusual star among them is not inconsistent with estimates of GC escapees in the halo of ~3\% \citep{Martell11}, noting that there are also two other marginal candidates with $\rm[Mg/Fe]\sim0$, $\rm[Al/Fe]\sim1$, and $\rm[Fe/H]=-0.6/-1.3$. We omit these additional stars from the discussion and plots because of the significantly lower S/N of their spectra.

While we consider a GC origin of 22593757-4648029 the most likely explanation for its non-typical abundance patterns, one may speculate on alternative reasons, e.g., mass transfer from a binary companion in the field. The so-called CEMP-s stars \citep[e.g.][]{Lucatello05} are a class of metal-poor stars believed to have been polluted with gas transferred from an AGB companion. These are characterised by enhancement in carbon and slow neutron-capture elements. The spectral range does not allow a stringent constraint on [C/Fe], but the upper limit $\rm[Y/Fe]<0.11$ excludes strong s-process enhancement. Further, CEMP stars are not characterised by low Mg abundances; no star in the sample presented by \citet[][including literature studies]{Allen12} has sub-solar [Mg/Fe]. According to \citet{Ventura11}, AGB stars with $\sim6M_{\odot}$ produce the most extreme Mg-Al-Si nucleosynthesis. If such a star was once a binary companion to our presumably old and low-mass field star, the system would have an unusual mass ratio. 

Low [Mg/Fe]-ratios are commonly found in dwarf spheroidal galaxies at this metallicity, but are then accompanied by similarly low ratios of other $\alpha$ elements, like Ca, Si, and Ti, with respect to Fe \citep[see e.g.][]{Koch08a,Kirby09}. No [Al/Fe] enhancement has been found (nor is it expected) in these systems.

\begin{figure}
\centering
\includegraphics[viewport=0cm 0cm 24cm 26.3cm, width=10.5cm,clip]{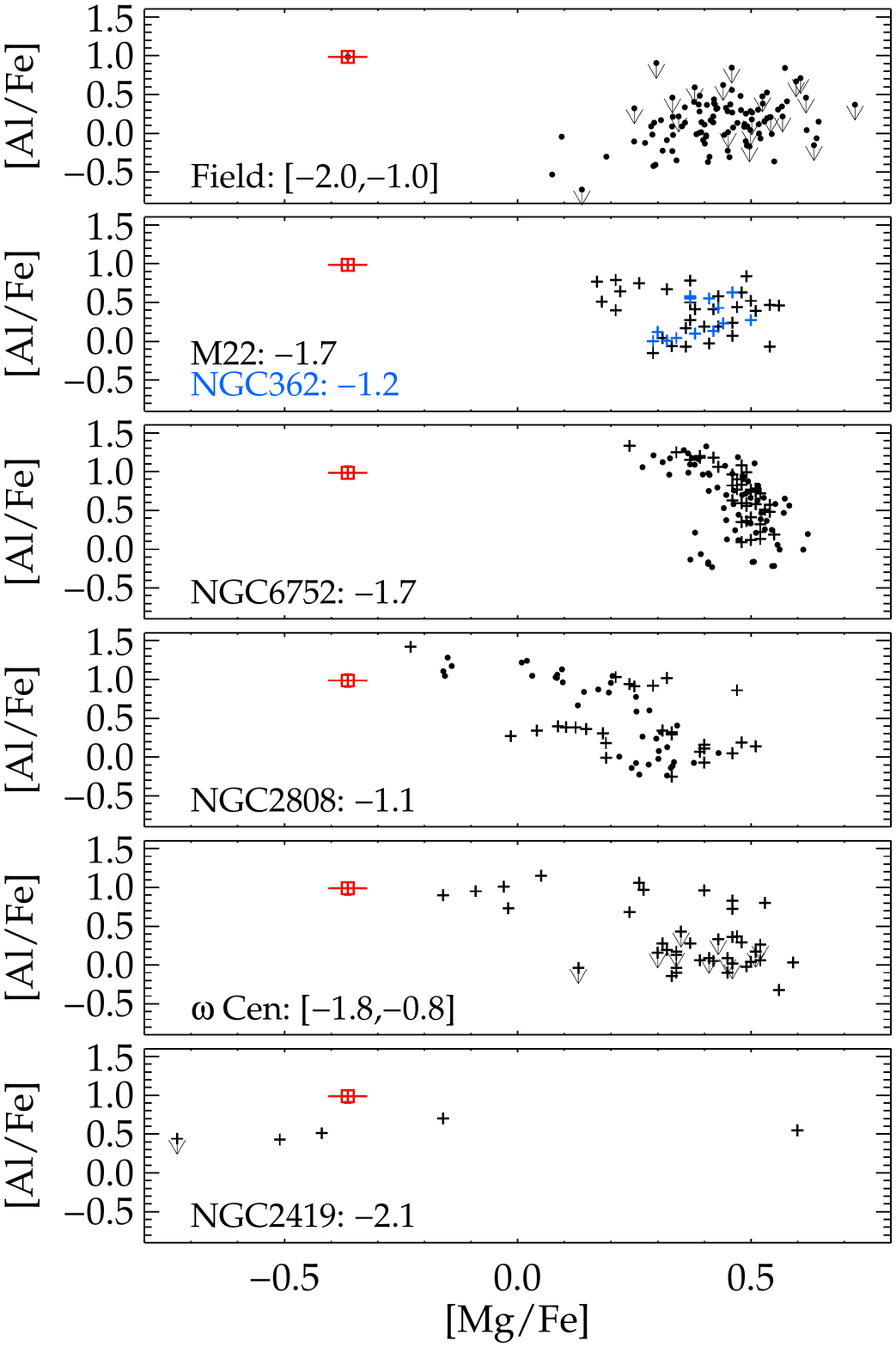}
\caption{Mg and Al abundances of normal field stars and five GCs compared to 22593757-4648029 (red square with error bars). The average metallicity is listed after the cluster name. Abundance data represented with plus signs are taken from \citeauthor{Worley10} (2010; NGC\,362), \citeauthor{Marino11} (\citeyear{Marino09}, 2011; M\,22), \citeauthor{Norris95} (1995; $\omega$ Cen), \citeauthor{Yong05} (2005; NGC\,6752), \citeauthor{Carretta06} (2006; NGC\,2808), \citeauthor{Carretta09b} (2009; NGC\,2808), and \citeauthor{Cohen12} (2012; NGC\,2419). GES recommended data are marked with bullets. Arrows indicate upper limits.}
\label{fig:hist}
\end{figure}

In Fig.\,\ref{fig:hist}, we compare the Mg and Al abundances of the field star to six different GCs. M\,22 and NGC\,362\footnote{Mg and Al abundance data for NGC\,362 were determined using spectra and stellar parameters obtained by \citet{Worley10}.}, like most GCs, display no striking Mg-Al anti-correlation and the lowest [Mg/Fe] values of these clusters are far from that of 22593757-4648029. NGC\,6752, NGC\,2808 and $\omega$ Cen all show strong evidence of multiplicity and internal He variations, having at least triple main sequences and extended horizontal branches \citep[e.g.][]{Piotto07,Milone13,King12}. As seen in Fig.\,\ref{fig:hist}, the [Al/Fe] ratios in the most highly polluted stars in these clusters are as high as that of 22593757-4648029, while the [Mg/Fe] ratios of stars in NGC\,2808 and $\omega$ Cen are almost as low. However, NGC\,2808 is too metal-rich to well match the metallicity of 22593757-4648029, making $\omega$ Cen the more likely parent GC. Unfortunately, there are surprisingly few Mg abundance measurements published for this otherwise well-studied GC. The fact that [Mg/Fe] ratios as low as $-0.36$ can be found in Galactic clusters is evidenced by the peculiar object NGC2419, where the record holder has [Mg/Fe]$\approx-1$ \citep{Mucciarelli12,Cohen12}. This cluster, however, is instead too metal-poor to be a plausible parent. 

It is easily realised that chemistry alone in not conclusive, because as many as 50 GCs have [Fe/H] within $\pm0.20$\,dex of the field star \citep[][2010 edition]{Harris96}  and most of them lack Mg and Al data. However, further insight can be obtained from the known kinematics of the star (see Table\,\ref{tab:star}) and about half of the candidate GCs with matching metallicity (D.\,Casetti\footnote{\url{http://www.astro.yale.edu/dana/gc.html}}). We proceed under the assumption that the progenitor cluster is still intact, while evidence  has also been found in the inner halo for disrupted GCs \citep{Bernard14}.

\begin{figure}
\centering
\includegraphics[width=7.8cm]{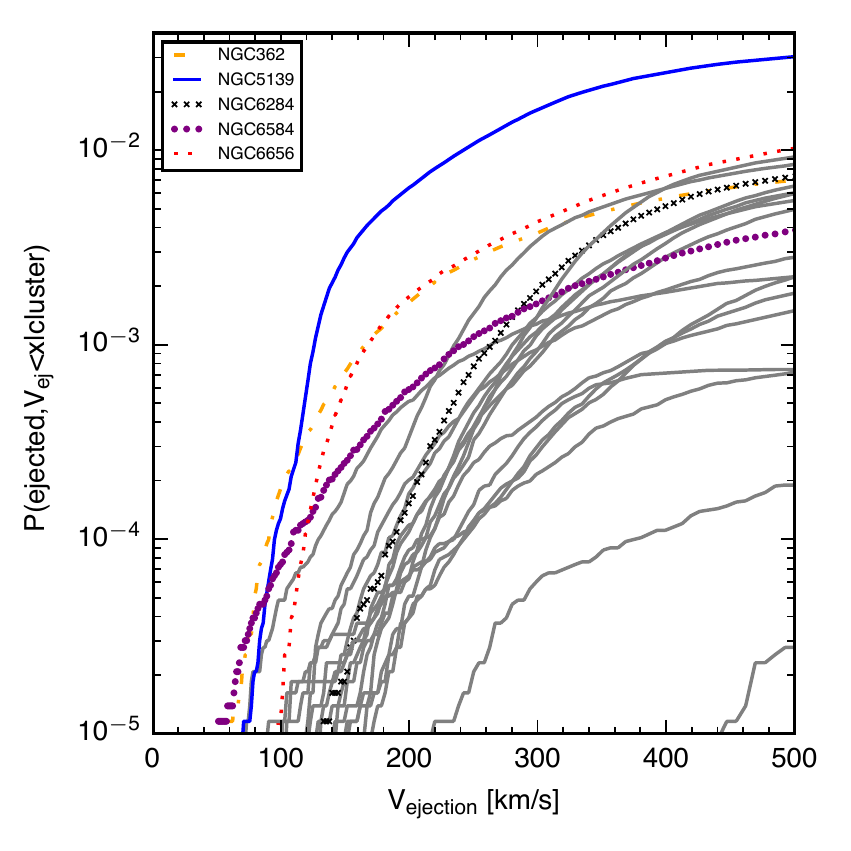}
\caption{The figure illustrates the cumulative probability that 22593757-4648029 was ejected from a given GC with a velocity lower than $V_{\rm ejection}$ within the last five billion years.}
\label{fig:probability}
\end{figure}

We assess the likelihood that the star was previously inside a known GC by integrating the current orbit of the star and the GC 5\,Gyr back in time in the Milky Way potential and look for close encounters. Following the work of e.g. \citet{Johnston99} and \citet{PriceWhelan13}, the energy of the star during a close encounter is computed as $E=\frac{1}{2}\delta V^2 + \Phi_{\rm GC}(\delta r)$ where $\delta V$ is the relative velocity of the star to the cluster and $\delta r$ is the distance from the cluster to the star. Encounters within $r_{\rm GC}$ and with $E<0$ are consistent with the star having been tidally stripped with a relative velocity smaller than the escape velocity, while encounters within $r_{\rm GC}$ and $E>0$ are consistent with the star having been ejected with a velocity of approximately $\sqrt{2\,E}$. We proceed to Monte-Carlo sample the proper motions, radial velocities, and distances of the star and of the clusters according to their error bars. Then, for each Monte-Carlo sample, we determine whether encounters occurred within the tidal radius and record the star's energy at the time. The fraction of Monte-Carlo samples with close encounters and $E< \frac{1}{2}V_{\rm ejection}^2$ therefore gives us a probability that a star was ejected from a given cluster with velocity $V<V_{\rm ejection}$ i.e. $P(\rm ejected, V<V_{\rm ejection}|\rm\,GC)$. The plot of this cumulative probability is shown in Figure\,\ref{fig:probability}. 

Evidently, given the current observational errors, it seems unlikely that the star has been tidally stripped from any cluster i.e. with small ejection velocity. But if we allow it to have been ejected at high velocity ($>100$\,km/s), which is possible in some  scenarios involving black holes and binary system interactions \citep{Gvaramadze09,Luetzgendorf12}, a handful of clusters could possibly have been associated with 22593757-4648029 in the past. Normalising the total probability to one, the top three posterior probabilities of ejections assuming $V_{\rm ejection}<$200\,km/s is NGC\,5139 ($\omega$ Cen) 44\%, NGC\,6656 (M\,22) 12\%, NGC\,362 12\% while for some clusters this probability is negligible, e.g. NGC\,5272, NGC\,5904, and NGC\,7006. More details of the orbit calculations are listed in Appendix A.

Interestingly, we note that $\omega$ Cen, which appeared most plausible already from the limited chemical information at hand is also favoured based on pure dynamical arguments. A more detailed chemical abundance analysis, including more elements, would help to verify this hypothesis. In particular, Mn and Cu abundances may help because a fraction of $\omega$ Cen stars are under-abundant in these elements compared to the field and other GCs \citep[][and references therein]{Cunha10,Romano07}. In a few years time, the Gaia mission will have greatly improved the accuracy and statistics of distance and proper-motion information for stars and star clusters. Combining this information with radial velocities and chemistry from ground-based telescopes, we will be able to connect stars like 22593757-4648029 to their parent GCs with greater confidence.  
 
\begin{acknowledgements}
This work was partly supported by the European Union FP7 programme through ERC grant number 320360, by the MICINN grant AYA2011-24704, by the ESF EUROCORES Program EuroGENESIS (MICINN  grant
EUI2009-04170), and by the Leverhulme Trust through grant RPG-2012-541. A.B., M.B., S.R., and D.R.  acknowledge the support from INAF and Ministero dell' Istruzione, dell' Universit\`a' e della Ricerca (MIUR) in the form of the grant "Premiale VLT 2012"  and PRIN MIUR 2010-2011 project "The Chemical and Dynamical Evolution of the Milky Way and Local Group Galaxies", prot. 2010LY5N2T. The results presented here benefit from discussions held during the Gaia-ESO workshops and conferences supported by the European Science Foundation (ESF) through the GREAT Research Network Programme. We further acknowledge support from The Swedish Research Council (T.B. funded by grant No. 621-2009-3911), the Swedish Space Board (U.H. and A.J.K), and the Australian Research Council (M.A. funded by grant FL110100012). K.L. acknowledge the European Union FP7-PEOPLE-2012-IEF grant No. 328098.
\end{acknowledgements}

\begin{appendix} 

\section{Orbit calculations}

The Milky Way potential that we used in the simulations is the analytical 3-component bulge-, disk-, and halo-potential used to model the orbits of satellites in \citet{Johnston95}, \citet{Dinescu99}, and \citet{Koposov10}. The potential of each cluster was approximated by a Plummer sphere using currently measured half-light radii and masses. The calculation of the tidal radii was done assuming constant rotation curve (e.g Milky Way enclosed mass is proportional to radius).

Since the true Galactic potential is not accurately known and the results of our simulations may depend significantly on the assumed parameters, we also run the simulations with masses of the disk and bulge perturbed by 10\% Gaussian variations and see how it affects the probabilities of the star being associated with one of the clusters. As an additional test, we also run our backwards orbit integration to 2\,Gyr instead of 5\,Gyr to see how it affects our results. While the individual probabilities for the clusters do change, the overall picture does not, e.g. $\omega$ Cen is always ranked within the top three.

 We further note some important limitations of our calculation: we ignore the dynamical friction of the GCs, which is expected to bring the GCs closer to the Galactic centre. We also ignore the fact that all the  GCs were likely to be bigger in the past and have tidally lost some of that mass, which reduces the absolute probabilities we find. Finally, we ignore secular changes in the Milky Way potential over time, which are expected although the Galaxy had a quiet recent accretion history.

\end{appendix}

\end{document}